\centerline{\bf CAN HST MEASURE THE MASS OF THE ISOLATED NEUTRON STAR}
\centerline{\bf RX J185635-3754 ?}
\vskip 0.5cm
\centerline{Bohdan Paczy\'nski}
\centerline{E-mail: bp@astro.princeton.edu}
\centerline{Princeton University Observatory, 124 Peyton Hall, Princeton, 
NJ 08544-1001, USA}
\vskip 0.5cm
\centerline{ABSTRACT}
\vskip 0.3cm

In June 2003 the isolated neutron star RX J185635-3754 will pass within 
$ \sim 0.3'' $ of a 26.5 mag star, changing its position by about 0.6 mas.
The displacement, caused by gravitational lensing, will be proportional to the
neutron star mass.  The total event duration will be approximately 1 year.

\vskip 0.2cm
\centerline{ - - - - - - - - }
\vskip 0.2cm

The possibility of measuring stellar masses using astrometric effects of
gravitational lensing were discussed by
Hog et al. (1995), Miyamoto \& Yoshii (1995), Paczy\'nski (1995, 1996, 1998),
Miralda-Escud\'e (1996), and Boden et al. (1997).  The recent detection of
a fast moving nearby neutron star RX J185635-3754 (Walter 2001) offers an
opportunity to implement this method in practice.  RX J185635-3754 is at 
the distance of 61 pc, and in June 2003 it will pass within 
$ {\rm \sim 0.3'' = 300 ~ mas } $ of the star marked as \# 115 in Fig. 1 of
Walter (2001).  The star \#115 has magnitude $ M_{F606W} = 26.5 $, the
neutron star has a parallax of 16 mas, and a proper motion of 
$ \dot \varphi = 332 $ mas/year.  

While the neutron star will be passing within $ \Delta \varphi \approx 300 $
mas of our line of sight towards the star \#115, the position of the latter
will be displaced by up to
$$
{\rm \delta \varphi = { \varphi _E ^2 \over \Delta \varphi } ,}
\eqno(1)
$$
where $ \varphi _E $ is the neutron star's Einstein ring radius (cf. Fig. 1
and eq. 4 of Paczy\'nski 1996).

Einstein ring radius is given with the eq. (1) of Paczy\'nski (1996):
$$
{\rm \varphi _E =  \left( { 4GM \over c^2 d_{\pi} } \right) ^{1/2} }
\eqno(2)
$$
where $ {\rm 1/d_{\pi} } $ is the parallax distance to the lens 
(RX J185635-3754) as measured with respect to the more distant source 
(star \#115):
$$
{\rm { 1 \over d_{\pi} } \equiv { 1 \over D_d } - { 1 \over D_s } }.
\eqno(3)
$$
Adopting $ {\rm D_d = 61 ~ pc} $, $ {\rm D_s/D_d \gg 1 } $, and
$ {\rm M \approx 1.4 ~ M_{\odot} } $ we obtain
$$
{\rm \varphi _E \approx 14 ~ mas }.
\eqno(4)
$$
By coincidence this Einstein ring radius is almost the same as the parallax 
of the neutron star.  The time scale for the lensing event will be
$$
{\rm \Delta t \equiv { \Delta \varphi \over \dot \varphi } \approx 1 ~ year .}
\eqno(5)
$$
The maximum displacement of the position of star \#115 due to lensing by
the neutron star will be approximately (cf. eq. 1):
$$
{\rm \delta \varphi \approx { (14 ~ mas )^2 \over 300 ~ mas } \approx 0.6 ~ 
mas }.
\eqno(6)
$$
This is a very small displacement, but it may be measurable with the new 
Advanced Camera for Surveys (ACS - http://www.stsci.edu/cgi-bin/acs),
which is to be mounted on the Hubble Space Telescope before the end of 2001.

The angles {\rm $ \varphi _E $, $ \Delta \varphi $, and $ \delta \varphi $
as given above, are only estimates.  However, when all these angles are 
measured the mass of the neutron star follows from the combination of
eqs. (1-3):
$$
{\rm { M \over M_{\odot} } =
{ c^2 d_{\pi} \over 4 G } ~ \delta \varphi ~ \Delta \varphi =
2.25 ~ { d_{\pi} \over 61 ~ pc } ~ { \delta \varphi \over 1 ~ mas } ~
{ \Delta \varphi \over 300 ~ mas } .}
\eqno(7)
$$

There are many precedents for the HST based determination of a parallax with
an accuracy of $ \sim 0.1 $ mas (e.g. McArthur et al. 2001), but those were
done for stars much brighter than either star \#115 or RX J185635-3754.  The
parallax of the faint neutron star was measured with an accuracy of $ \sim 2 $
mas with the Wide Field Planetary Camera 2 (Walter, 2001).  The HST astrometry
should be more accurate with the Advanced Camera for Surveys, which will become
operational by early 2002, with plenty of time to make the preparations for the
lensing event of June 2003.  Such a measurement may provide a novel check on
the estimates made by Pons et al.  (2001)

This paper is posted on astro-ph only, and it will not be submitted to any
paper journal.  The author welcomes all critical comments by the readers, 
in particular about missing references.  It is a pleasure to acknowledge 
the support from the NSF grant AST-9820314.

\vskip 0.5cm

\centerline{\bf REFERENCES}

\vskip 0.3cm

Boden, A. F., Shao, M., \& Van Buren, D. 1997, ApJ, 502, 538

Hog, E., Novikov, I. D., \& Polnarev, A. G. 1995, A\&A, 294, 287 

McArthur, B. E. et al. 2001, astro-ph/0107026

Miralda-Escud\'e, J. 1996, astro-ph/9605138 = ApJ, 470, L113

Miyamoto, M., \& Yoshii, Y, 1995, AJ, 110, 1427

Paczy\'nski, B. 1995, astro-ph/9504099 = AcA, 45, 345

Paczy\'nski, B. 1996, astro-ph/9606060 = AcA, 46, 291

Paczy\'nski, B. 1998, astro-ph/9708155 = ApJ, 494, L23

Pons, J. A. et al. 2001, astro-ph/0107404

Walter, F. M. 2001, astro-ph/0009031 = ApJ, 549, 433

\vfill \end \bye